\documentstyle[aps]{revtex}
\begin{document}
\title{Dirty Hubbard rings:
renormalization group and exact diagonalization studies}
\author{Hiroyuki Mori and Minoru Hamada}
\address{Department of Materials Science,
Faculty of Science, Hiroshima University,\\
Higashi-Hiroshima 739, Japan}
\maketitle
%
%
\begin{abstract}
We studied mesoscopic Hubbard rings with impurities
using the renormalization group (RG) technique
and the exact diagonalizations.
The exact diagonalization calculations showed that
the charge stiffness normalized by the value
of clean system has a peak as a function of
electron-electron interactions.
Previous RG analysis has succeeded to show the
enhancement of the stiffness at weak interactions, but does not
show the peak behavior.
We derived RG equations with $4k_F$ impurity scattering term,
which was ignored in the previous studies, and
reproduced the overall
behavior of the exact diagonalization results.
\end{abstract}
\tighten
%
%
\section{Introduction}
Low dimensional systems often
show interesting quantum phenomena,
and there particle-particle interactions and impurities
sometimes play crucial roles. Mesoscopic rings are known to be
one of the good examples and have been studied
over the last few years in particular.
When an external flux is applied through a mesoscopic ring,
the current flows along the ring and the magnitude
and the direction of the current changes
as a function of the applied flux. The current
flows in equilibrium and is therefore called persistent current.
The presence of persistent current was first predicted by
Byers and Yang, and B\"{u}ttiker {\it et al.}\cite{byers},
and observed in several recent experiments\cite{levy}.
The magnitude of the observed persistent current
is, however, much larger than the theoretical value
calculated for non-interacting electrons
with impurities\cite{cheung},
and therefore we need to take account of other factors.
The factors, other than impurities,
that might affect the persistent
current are, for example, particle-particle interaction,
periodic potential, spin degree of freedom, and multichannels.
If the system has only one of these, the situation is
rather simple and we believe we already well understand it.
Impurities would reduce the persistent current\cite{cheung}.
Interactions would not change
the persistent current\cite{muller,loss}.
Periodic potentials would change
the shape of the persistent current
as a function of the applied flux,
but would not change its magnitude
very much\cite{trivedi}.
Spin degree of freedom or multichannels just give
some numerical factor to
the magnitude of the persistent current.

Now let us consider the system that
has two or more of the above factors.
The role of each factor could be different in this case.
In the presence of particle-particle interactions {\em and}
impurities, for example, the persistent current decreases
as the interactions increase\cite{muller,romer},
while the interaction is an irrelevant parameter when it
exists with no other factors.
In general\cite{muller},
interaction becomes relevant if it coexists with
translationally noninvariant factors.
As this example shows, the situations
when two or more factors coexist are usually complicated
and therefore interesting.
In this paper we focus on the system
with impurities, interactions,
periodic potentials, and spins.
The most typical and maybe the simplest
model that contains all of them would be dirty Hubbard ring.
Dirty Hubbard ring has been recently studied
analytically\cite{giamarchi1,romer} and
numerically\cite{deng,kato2}.
Giamarchi and Shastry\cite{giamarchi1} showed
using the renormalization group (RG) analysis that
repulsive interactions between particles would enhance
the persistent current while they would suppress
the persistent current
in spinless fermion systems\cite{bouzerar,kato1,mori}.
The persistent current does not
continue to be enhanced, however,
as we increase the interactions further.
The exact diagonalization
calculations\cite{deng,kato2} show
that the persistent current has a peak
at a characteristic interaction strength
and begins to be suppressed
by the interaction beyond that point.
Two origins are expected to give the suppression.
One is periodic potential.
Even in the absence of impurities, interactions would suppress
the persistent current on a lattice
and the suppression is more emphasized
as the particle density approaches to the half filling.
This effect would exist in the dirty systems.
The second origin that would suppress the persistent current
would be the $4k_F$ Fourier component of impurity scatterings.
For strong electron-electron interactions,
Hubbard model approaches to
a spinless fermion model,
and the $4k_F$ scatterings in Hubbard model
corresponds to the $2k_F$ scatterings
in the spinless fermion model.
Since it is known\cite{bouzerar,kato1,mori}
that the $2k_F$ scatterings in the spinless fermion model
would suppress the persistent current,
we expect the $4k_F$ scatterings
would also suppress the persistent current of Hubbard model.
The previous RG equations do not contain
the $4k_F$ scattering term
and are not complete to reproduce the numerical results.
We will derive RG equations
taking account of the $4k_F$ scatterings
and will show the peak behavior of the persistent current.

The paper is organized as follows. In Sec. 2 we show the
results of our exact diagonalization calculations.
RG analysis is given in Sec. 3. Section 4 is devoted to
summary of this study.
%
%
\section{Numerical results}
We first show our exact diagonalization results, which
are basically the same calculations
as the previous ones\cite{deng,kato2}.
We used the modified Lanzcos method\cite{dagotto}
to calculate the ground state energy $E$ of Hubbard ring:
\begin{equation}
H=-t\sum_{<ij>,\sigma }e^{i2\pi\phi /\phi_0 L}
c_{i\sigma }^{\dagger }c_{j\sigma }+c.c.
+U\sum_{i}n_{i\uparrow }n_{i\downarrow }
+\sum_{i\sigma}V_{imp}(i)n_{i\sigma}
\end{equation}
as a function of the applied flux $\phi$.
We assumed the uniform probability
for the impurity potential $V_{imp}$ between $-W$ and $W$.
The impurity average of the energy $E$ was taken
over 512 to 3072 realizations.
The persistent current $I(\phi)$ is
given by the first derivative
of the averaged energy with respect to the flux $\phi$,
and the charge stiffness $D_c$ is proprtional to
the second derivative of the energy.
Since it it easier to calculate
the charge stiffness than to calculate
the persistent current in the RG analysis,
and since the charge stiffness is a measure
of the persistent current\cite{giamarchi1,mori},
we here show the results
of the charge stiffness calculations.
In the calculations we always chose even number of
particles per spin. The reason is the following.
For simplicity, let us consider free electrons
with the numbers of up and down spins, $N_{\uparrow}$ and
$N_{\downarrow}$, being equal.
The ground state energy of the free electrons changes
as we change the particle number per spin,
$N_{\sigma}$, from even to odd,
and it has the relation $E_{even}(\phi)=E_{odd}(\phi+\phi_0/2)$
where $\phi_0$ is the flux quanta.
If we define the charge stiffness by
$D_c=(L/2)\partial^2E(\phi)/\partial\phi^2|_{\phi=0}$,
the stiffness as a function of the particle density,
does not change smoothly when $N_{\sigma}$
moves between even and odd.
We can avoid this problem if we take the second derivative
at $\phi=0$ when $N_{\sigma}$ is odd,
and at $\phi=\phi_0/2$  when $N_{\sigma}$ is even.
This causes another problem, however.
In $U\rightarrow\infty$ limit
Hubbard model becomes the free spinless fermion model, where
the particle number is always even
since $N_{\uparrow}=N_{\downarrow}$.
Therefore we should take the derivative of
the ground state energy
at $\phi=\phi_0/2$ in $U\rightarrow\infty$
limit although the derivative has to be taken at $\phi=0$
at small $U$.
If we choose $N_{\sigma}$ to be even,
the system is always in the `even number` sector in both limits
of $U=0$ and $U\rightarrow\infty$, and we do not have to change
the definition of the stiffness. For this reason,
we chose even number of particles per spin and used the definition
$D_c=(L/2)\partial^2E(\phi)/\partial\phi^2|_{\phi=\phi_0/2}$
in the following exact diagonalization calculations.

Figure (\ref{exact1}) shows the charge stiffness
$D_c$ as a function of
$U/t$ in the system of $L$=6 and $N_{\uparrow}=N_{\downarrow}=2$.
As observed in the previous exact diagonalization
studies\cite{deng,kato2},
it has a peak, and the peak position shifts towards
higher $U/t$ for the stronger impurity potential.
The enhancement at small $U$ is due to the impurity scatterings,
since we know the charge stiffness
in the clean systems, $D_c(W=0)$,
decreases monotonically as $U$ becomes larger,
If we normalize the stiffness by $D_c(W=0)$,
the peak still exists and
the peak positions are located almost in the same place
for the different impurity strengths (Fig. (\ref{exact2})).
Since the normalization effectively eliminates the
periodic potential contribution, Fig. (\ref{exact2}) suggests
that the reduction of the charge stiffness at large $U$
is not only due to the periodic potentials but also
due to the impurities.
In the next section we will try to reproduce this behavior
of the charge stiffness using the RG technique.
%
%
\section{RG calculations}
One dimensional Hubbard model with impurities can be
written in the boson representation by the following Lagrangian:
\begin{equation}
L=L_{\rho}+L_{\sigma}+L_{g}+L_{2k_F}+L_{4k_F},
\label{lagrangian}
\end{equation}
where
\begin{eqnarray*}
L_{\nu}&=&\frac{1}{2\pi K_{\nu}}\int dx \{u_{\nu}^{-1}
(\partial_{\tau}\phi_{\nu}(x,\tau ))^2
+u_{\nu}(\partial_{x}\phi_{\nu}(x,\tau ))^2\},\\
L_{g}&=&\frac{2g_{1\perp}}{(2\pi\alpha)^2}\int dx
\cos (\sqrt{8}\phi_{\sigma}(x,\tau)),\\
L_{2k_F}&=&\frac{1}{\pi\alpha}
\int dx \xi_2 (x)e^{i\sqrt{2}\phi_{\rho}(x,\tau)+2k_Fx}
\cos (\sqrt{2}\phi_{\sigma}
(x,\tau)) +h.c,\\
L_{4k_F}&=&\frac{1}{\pi\alpha}
\int dx \xi_4 (x)e^{i\sqrt{8}\phi_{\rho}(x,\tau)+4k_Fx}
+h.c,
\end{eqnarray*}
$\nu=\rho, \sigma$, and $\alpha$ is the lattice constant.
The last two terms in Eq. (\ref{lagrangian})
represent the $2k_F$ and $4k_F$ Fourier components of
impurity scatterings, respectively. The parameter $g_{1\perp}$
is equal to $U\alpha$,
where $U$ is the on-site repulsion of Hubbard model.
We assume gaussian distribution
for the random fields $\xi_2$ and
$\xi_4$:
$P(\xi_2)=N_2\exp (-\int |\xi_2(x)|^2dx/W_2)$
and
$P(\xi_4)=N_4\exp (-\int |\xi_4(x)|^2dx/W_4)$,
where $N_2$ and $N_4$ are the normalization factors.
We did not include the forward scattering due to impurities
because it can be eliminated
by redefining the field $\phi_{\rho}$\cite{giamarchi2}.

With help of the replica trick,
we can integrate out the random fields\cite{giamarchi2}
and the final form of the action is
\begin{equation}
S=S_{\rho}+S_{\sigma}+S_{g}+S_{2k_F}+S_{4k_F},
\end{equation}
where
\begin{eqnarray*}
S_{\nu}&=&\sum_i\frac{1}{2\pi K_{\nu}}
\int dxd\tau \{u_{\nu}^{-1}
(\partial_{\tau}\phi_{\nu}^i(x,\tau ))^2
+u_{\nu}(\partial_{x}\phi_{\nu}^i(x,\tau ))^2\},\\
S_{g}&=&\sum_i\frac{2g_{1\perp}}{(2\pi\alpha)^2}\int dxd\tau
\cos (\sqrt{8}\phi_{\sigma}^i(x,\tau)),\\
S_{2k_F}&=&\sum_{ij}\frac{W_2}{(\pi\alpha)^2}\int dxd\tau d\tau '
\cos (\sqrt{2}\phi_{\sigma}^i(x,\tau ))
\cos (\sqrt{2}\phi_{\sigma}^j(x,\tau '))
\cos (\sqrt{2}(\phi_{\rho}^i(x,\tau )
-\phi_{\rho}^j(x,\tau ')),\\
S_{4k_F}&=&\sum_{ij}\frac{W_4}{(\pi\alpha)^2}\int dxd\tau d\tau '
\cos (\sqrt{8}(\phi_{\rho}^i(x,\tau )-\phi_{\rho}^j(x,\tau ')),
\end{eqnarray*}
where $i$ and $j$ are the replica indices.
Now we are ready to derive the RG equations.
Up to the lowest order of $g_{1\perp}, W_2$ and $W_4$,
the RG equations are obtained in the following form:
\begin{eqnarray}
\frac{dK_{\rho}(l)}{dl}&=&-\frac{1}{2}
\left(\frac{K_{\rho}^2u_{\rho}}{u_{\sigma}}\right)
(\Delta_2(l)+\Delta_4(l)),\nonumber\\
\frac{du_{\rho}(l)}{dl}&=&-\frac{1}{2}
\left(\frac{K_{\rho}u_{\rho}^2}{u_{\sigma}}\right)
(\Delta_2(l)+\Delta_4(l)),\nonumber\\
\frac{dK_{\sigma}(l)}{dl}&=&-\frac{1}{2}
K_{\sigma}^2(\Delta_2(l)+y(l)^2),\nonumber\\
\frac{du_{\sigma}(l)}{dl}&=&-\frac{1}{2}
K_{\sigma}u_{\sigma}\Delta_2(l),
\label{RG}\\
\frac{dy(l)}{dl}&=&(2-2K_{\sigma}(l))y(l)-W_2(l),\nonumber\\
\frac{\Delta_2(l)}{dl}&=&(3-K_{\rho}(l)-K_{\sigma}(l)-y(l))
\Delta_2(l),\nonumber\\
\frac{\Delta_4(l)}{dl}&=&(3-4K_{\rho}(l))
\Delta_4(l),\nonumber
\end{eqnarray}
where $y=g_{1\perp}/\pi u_{\sigma}$,
$\Delta_2=(2W_2\alpha/\pi u_{\sigma}^2)%
(u_{\sigma}/u_{\rho})^{K_{\rho}}$,
and $\Delta_4=8W_4\alpha u_{\sigma}/\pi u_{\rho}^3$.
If we ignore the $4k_F$ scatterings, i.e. if we set $W_4=0$,
the RG equations reduce to the ones
derived by Giamarch and Schulz\cite{giamarchi2}.

We solved the Bethe ansatz equations of
clean Hubbard ring\cite{lieb,shastry}
to determine the initial values
for the integration of the RG equations.
{}From the Bethe ansatz equations we obtain the stiffness and
the susceptibility of charge and spin sectors.
The parameters $K$'s and $u$'s can then be calculated
from the following relations\cite{shastry,romer}:
\begin{eqnarray}
K_{\rho}&=&\pi\frac{\rho}{2}\sqrt{2D_{\rho}\chi_{\rho}},
\nonumber\\
u_{\rho}&=&\frac{2}{\rho}\sqrt{\frac{D_{\rho}}{2\chi_{\rho}}},
\nonumber\\
\label{Ku}
K_{\sigma}&=&\pi\rho\sqrt{2D_{\sigma}\chi_{\sigma}},\\
u_{\sigma}&=&\frac{4}{\rho}
\sqrt{\frac{D_{\sigma}}{2\chi_{\sigma}}},\nonumber
\end{eqnarray}
where  $D_{\nu}$ and $\chi_{\nu}$ are, respectively, the
stiffness and the susceptibility
of the $\nu(=\rho, \sigma)$ sector
and $\rho$ is the particle number density.
Note that the first two equations in Eq. (\ref{Ku}) are exact,
while the last two are approximate and valid
when $U/t$ is small\cite{shastry,romer}.

Using the obtained parameters
as the initial value of the RG equations,
we integrated the equations numerically and stopped when $l$
reached $\log (L/\alpha)$.
Figure (\ref{w2}) shows the normalized charge stiffness
$D_c/D_c(W=0)$
calculated with $W_4=0$. As obtained by Giamarchi and Shastry
\cite{giamarchi1}, the stiffness grows
as the interactions increase,
and never drops down. When $W_4$ is finite,
however, the normalized
charge stiffness has a peak as seen in Fig. (\ref{w4}).
The peak stays almost in the same position
for the different impurity potential strengths,
and this feature is also observed
in the exact diagonalization calculations.
We therefore believe the RG equations (\ref{RG}) are complete
to reproduce the qualitative feature of
the exact diagonalization calculations.
%
%
\section{Summary}
The previous RG calculations
that contain the $2k_F$ impurity scatterings
succeeded to show the enhancement
of the charge stiffness by the interaction $U$,
which is an opposite feature
to the spinless fermion systems where
the interactions would suppress the stiffness.
As we increase $U$ further, however,
the exact diagonalization studies show
the stiffness starts to decrease
making a peak at a characteristic $U$.
The peak behavior still exists even when we plot the normalized
stiffness $D_c(W)/D_c(W=0)$ as a function of $U$.
This feature cannot be obtained from the previous RG equations
and we need to take account of
the missing term in their calculations.

We derived the RG equations for the dirty Hubbard rings
with the $4k_F$ as well as
the $2k_F$ impurity scatterings being included,
and calculated the charge stiffness $D_c$ for various
impurity strengths $W$ and electron-electron interactions $U$.
At $U=\infty$ the Hubbard model is equivalent to
the free spinless fermion
model, where fermi wave vector is twice as large as
the one of the original Hubbard model.
Therefore the $4k_F$ scatterings become dominant as we approach
$U/t\rightarrow\infty$ limit.
Since the RG and the numerical calculations
for the spinless fermions showed
interactions would suppress
the charge stiffness in dirty systems,
we can expect that the RG equations with the $4k_F$ term should
give the decreasing behavior
of the charge stiffness with increasing $U/t$.
We numerically integrated the RG equations and
the normalized stiffness of the results makes
a peak at a certain $U/t$, while it grows monotonically
without the $4k_F$ term.
It is also found that the peak position
does not change very much for different impurity strengths.
These results are consistent to the numerical analysis.

If we ignore the $4k_F$ term in the RG analysis,
the persistent current becomes close to the impurity-free case
at large $U$.
In real systems, however, the $4k_F$ scatterings do exist and
the persistent current cannot reach the impurity-free value.
Therefore, one-dimensional Hubbard rings may not be sufficient
to explain the large persistent current observed in
the experiments\cite{levy},
and we should take account of other factors such
as multichannels.

Let us close this paper by commenting
on the Coulomb interaction cases.
Since arbitrary weak Coulomb interaction in one-dimensional
electron systems makes the
$4k_F$ density correlation dominant and drives the system
to a Wigner crystal\cite{schulz},
the $4k_F$ component of impurity scatterings
always plays central roles,
and the persistent current and the charge stiffness
would always be suppressed
by the Coulomb interactions as a consequence.
The exact diagonalization calculations\cite{kato2} actually
show that the Coulomb interactions would suppress
the persistent current unless the impurity potentials
are too strong\cite{comment}.
\acknowledgments
The authors would like to acknowledge
H.Kato for giving us the details of
their numerical data.
H.M. is supported by the Sasakawa Scientific Research Grant
from the Japan Science Society and
M.H. is supported by the Japan Society
for the Promotion of Science.
%
%

\begin{figure}
\caption{
Charge stiffness vs U/t for $W=0.1, 0.5,$ and $1.0$ from top
to bottom. The system size is $L=8$. The particle
numbers are $N_{\uparrow}=N_{\downarrow}=2$.
}
\label{exact1}
\end{figure}
\begin{figure}
\caption{
Normalized charge stiffness vs U/t.
The parameters are the same as in Fig. (1)
}
\label{exact2}
\end{figure}
\begin{figure}
\caption{
Normalized charge stiffness vs $U/t$ when $\Delta_4=0$.
The system size and the particle number per spin are
$L=100$ and $N_{\uparrow}=N_{\downarrow}=22$ respectively.
The initial value of
$\Delta_2$ is set to be $1, 30, 60\ (\times 10^{-4})$
from top to bottom.}
\label{w2}
\end{figure}
\begin{figure}
\caption{
The normalized charge stiffness vs
$U/t$ when $\Delta_4=4\Delta_2$.
The system size, the particle number, and $\Delta_2$
are the same as in Fig. (3)}
\label{w4}
\end{figure}

\begin{references}
\bibitem{comment}When the impurity potentials
are strong enough to localize the particles,
interactions could enhance the current.
See, for example, Ref. \cite{mori}.
\end{references}
\end{document}